\documentclass{article}

\usepackage{arxiv}

\usepackage[utf8]{inputenc} 
\usepackage[T1]{fontenc}    
\usepackage{hyperref}       
\usepackage{url}            
\usepackage{booktabs}       
\usepackage{amsfonts}       
\usepackage{nicefrac}       
\usepackage{microtype}      
\usepackage{lipsum}		
\usepackage{graphicx}
\usepackage{natbib}
\usepackage{doi}
\usepackage{amsmath}

\title{Recovering Sub-threshold S-wave Arrivals in Deep Learning Phase Pickers via Shape-Aware Loss}

\author{ \href{https://orcid.org/0000-0001-6805-2817}{\includegraphics[scale=0.06]{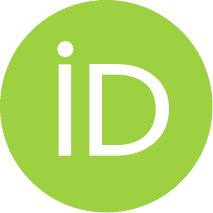}\hspace{1mm}Chun-Ming Huang}\\
	Department of Geosciences\\
	National Taiwan University\\
	Taipei, Taiwan \\
	\texttt{jimmy60504@gmail.com} \\
	\And
    \hspace{1mm}Li-Heng Chang\\
	Department of Geosciences\\
	National Taiwan University\\
	Taipei, Taiwan \\
	\texttt{qwert159784623@gmail.com} \\
    \And
    \hspace{1mm}I-Hsin Chang\\
	Department of Geosciences\\
	National Taiwan University\\
	Taipei, Taiwan \\
	\texttt{atetgod000@gmail.com} \\
    \And
	\href{https://orcid.org/0000-0002-5492-1986}{\includegraphics[scale=0.06]{orcid.pdf}\hspace{1mm}An-Sheng Lee} \\
	Marine Core Research Institute\\
	Kochi University\\
	Kochi, Japan \\
	\texttt{aslee@kochi-u.ac.jp} \\
    \And
	\href{https://orcid.org/0000-0003-2282-9218}{\includegraphics[scale=0.06]{orcid.pdf}\hspace{1mm}Hao Kuo-Chen} \\
	Department of Geosciences\\
	National Taiwan University\\
	Taipei, Taiwan \\
	\texttt{kuochenhao@ntu.edu.tw} 
}

\renewcommand{\shorttitle}{\textit{arXiv} Template}

\hypersetup{
pdftitle={A template for the arxiv style},
pdfsubject={q-bio.NC, q-bio.QM},
pdfauthor={David S.~Hippocampus, Elias D.~Striatum},
pdfkeywords={First keyword, Second keyword, More},
}

\begin{document}
\maketitle

\begin{abstract}
Deep learning has transformed seismic phase picking, but a systematic failure mode persists: for some S-wave arrivals that appear unambiguous to human analysts, the model produces only a distorted peak trapped below the detection threshold, even as the P-wave prediction on the same record appears flawless. By examining training dynamics and loss landscape geometry, we diagnose this amplitude suppression as an optimization trap arising from three interacting factors. Temporal uncertainty in S-wave arrivals, CNN bias toward amplitude boundaries, and the inability of pointwise loss to provide lateral corrective forces combine to create the trap. The diagnosis reveals that phase arrival labels are structured shapes rather than independent probability estimates, requiring training objectives that preserve coherence. We formalize this as the shape-then-align strategy and validate it through a conditional GAN proof of concept, recovering previously sub-threshold signals and achieving a 64\% increase in effective S-phase detections. Beyond this implementation, the loss landscape visualization and numerical simulation techniques we introduce provide a general methodology for analyzing how label designs and loss functions interact with temporal uncertainty, transforming these choices from trial-and-error into principled analysis.

\end{abstract}


\section{Introduction}

Seismic phase picking requires precise identification of seismic wave arrival times. Deep learning, particularly U-Net-based PhaseNet \citep{zhu2018phasenet-82a}, transformed this field by representing discrete arrival times as truncated Gaussian probability distributions. A systematic failure mode persists, however, where certain S-wave arrivals that appear obvious to human analysts remain sub-threshold in model predictions, even when P-waves on the same record are detected flawlessly.

This phenomenon, which we term amplitude suppression, has been widely observed but barely explained. The typical response has been to refine architectures \citep{mousavi2020earthquake-3d2, liao2021arru-ddf}, augment datasets \citep{li2024conseisgen-cad, wang2021seismogen-d95}, modify training objectives \citep{park2025reducing-831}, or bypass Gaussian templates entirely through direct regression \citep{williams2023deep}. These approaches treat suppression as a symptom to be worked around rather than a mechanism to be understood.

We take a different approach and focus on diagnosing the root cause. Training dynamics and loss landscape geometry are the two windows into gradient descent, the optimization engine underlying all deep learning. By examining how predictions evolve during training and how the loss function shapes the optimization surface the model navigates, we identify an optimization trap arising from three interacting factors. S-wave onsets are temporally uncertain, separated from subsequent high-amplitude boundaries by gaps that vary with epicentral distance. CNN correlation bias anchors predictions to these sharp boundaries rather than subtle onsets. And pointwise BCE loss provides only vertical gradients, lacking lateral forces to correct the resulting temporal misalignment.

The diagnosis points directly to a solution. Phase arrival labels are Gaussian templates designed by convention, but pointwise loss fragments them into independent decisions. Shape-aware training can restore coherence. We formalize this as the shape-then-align strategy, where geometric templates must stabilize before temporal alignment can succeed. As proof of concept, we implement a conditional GAN framework \citep{mirza2014conditional-023} where a discriminator enforces shape constraints alongside conventional BCE training.

Testing on the INSTANCE dataset \citep{michelini2021instance-b0c}, our approach recovers sub-threshold S-wave arrivals, achieving a 64\% increase in effective detections. The contribution is twofold. First, we provide a diagnostic framework built on loss landscape and training dynamics analysis, explaining why pointwise optimization fails under temporal uncertainty. Second, we validate that shape-aware loss can break this trap.

\section{Observations}

Before diagnosing the mechanism, we first document the phenomenon through direct observation. Training histories on individual seismic records reveal the dynamics of amplitude suppression in action.

\begin{figure}
    \centering
    \includegraphics[width=0.6\linewidth]{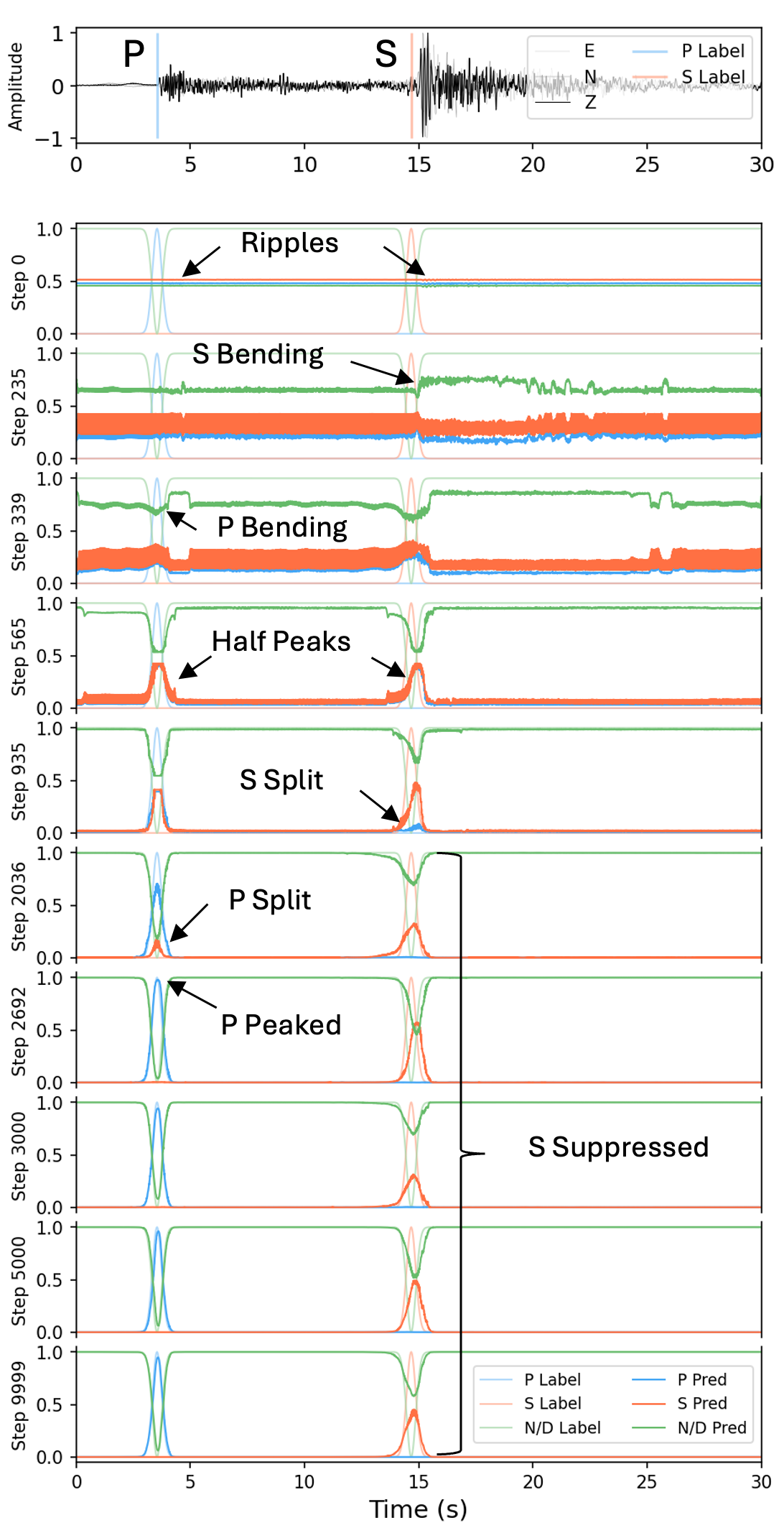}
\caption{\textbf{The dynamic process of amplitude suppression.}
Color coding: P-wave (blue), S-wave (orange), Noise (green); light/dark colors show labels/predictions. This figure visualizes key training steps from a conventional supervised model. The model exhibits a specific learning sequence: initially producing right-flanked responses at regions of sharpest amplitude change, then learning left flanks to form half-Gaussian curves with plateaus, and finally attempting to differentiate P/S features and grow peak amplitudes. However, even in late training stages, the S-phase peak remains trapped at $\sim$0.5 amplitude, oscillating in this suboptimal state. See Videos 1 and 2 in Data and Resources for the training animations (10,000 and 100,000 steps, respectively).\\[6pt]
\textbf{Alt-text:} Two-part figure. Top panel shows 3-channel seismic waveform from a regional event with vertical lines marking P and S arrival times. Bottom section contains 10 panels showing model prediction evolution from step 0 to 9999, displaying probability curves for P, S, and Noise channels.}
\label{fig:amplitude_suppression}
\end{figure}

The conventional PhaseNet model \citep{zhu2018phasenet-82a} trained on the INSTANCE dataset \citep{michelini2021instance-b0c} on a representative event shows a complex and divergent process for P- and S-waves (Fig. \ref{fig:amplitude_suppression}). Initially, both predictions evolve similarly, forming a half-peak platform. A clear divergence occurs after the model begins to separate their features. While the P-phase prediction successfully grows in amplitude, the S-phase's development is arrested as it approaches an amplitude of approximately 0.5, remaining trapped in amplitude suppression.

\begin{figure}
    \centering
    \includegraphics[width=1\linewidth]{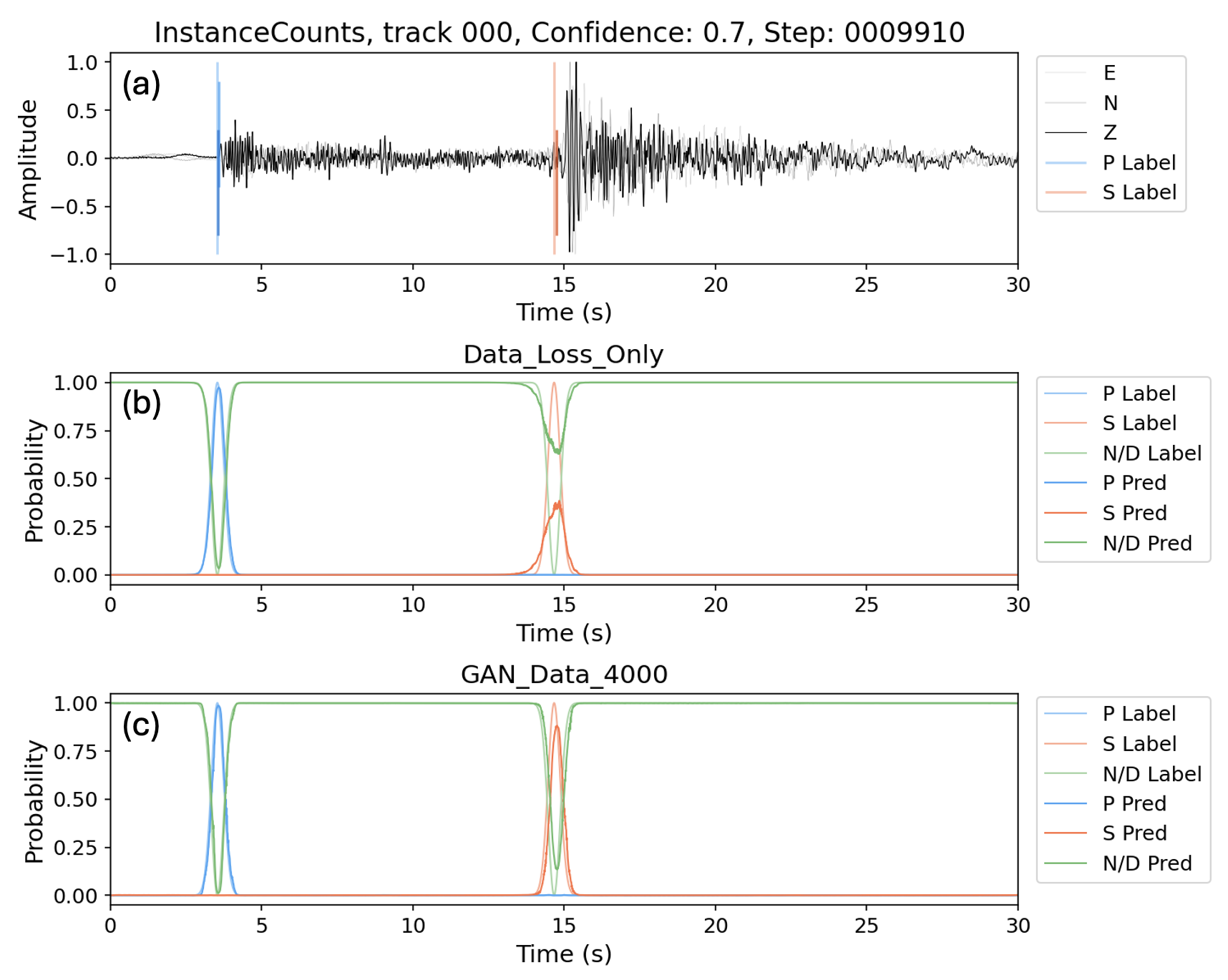}
\caption{\textbf{Visual diagnosis and correction of amplitude suppression.}
Color coding: P-wave (blue), S-wave (orange), Noise (green); light/dark colors show labels/predictions. (a) Representative waveform with phase arrival markers: longest line (reference label), upper short line (conventional prediction, panel b), lower short line (our framework, panel c). Note temporal delay between S-wave onset and high-amplitude wave packet. (b) Conventional method exhibits two suppression symptoms: peak amplitude suppressed at 0.5 (below 0.7 threshold) and temporal position shifted rightward toward high-amplitude boundary. (c) Our framework overcomes suppression: peak exceeds 0.7 threshold and temporal position calibrated within ±0.1 s accuracy. See Videos 1 and 2 in Data and Resources for the training animations (10,000 and 100,000 steps, respectively).\\[6pt]
\textbf{Alt-text:} Three vertically stacked panels. Top panel shows 3-channel seismic waveform with vertical lines marking phase arrivals. Middle and bottom panels show probability curves for P, S, and Noise channels from conventional and proposed methods respectively.}
    \label{fig:compare_runs}
\end{figure}

\begin{figure}
    \centering
    \includegraphics[width=1\linewidth]{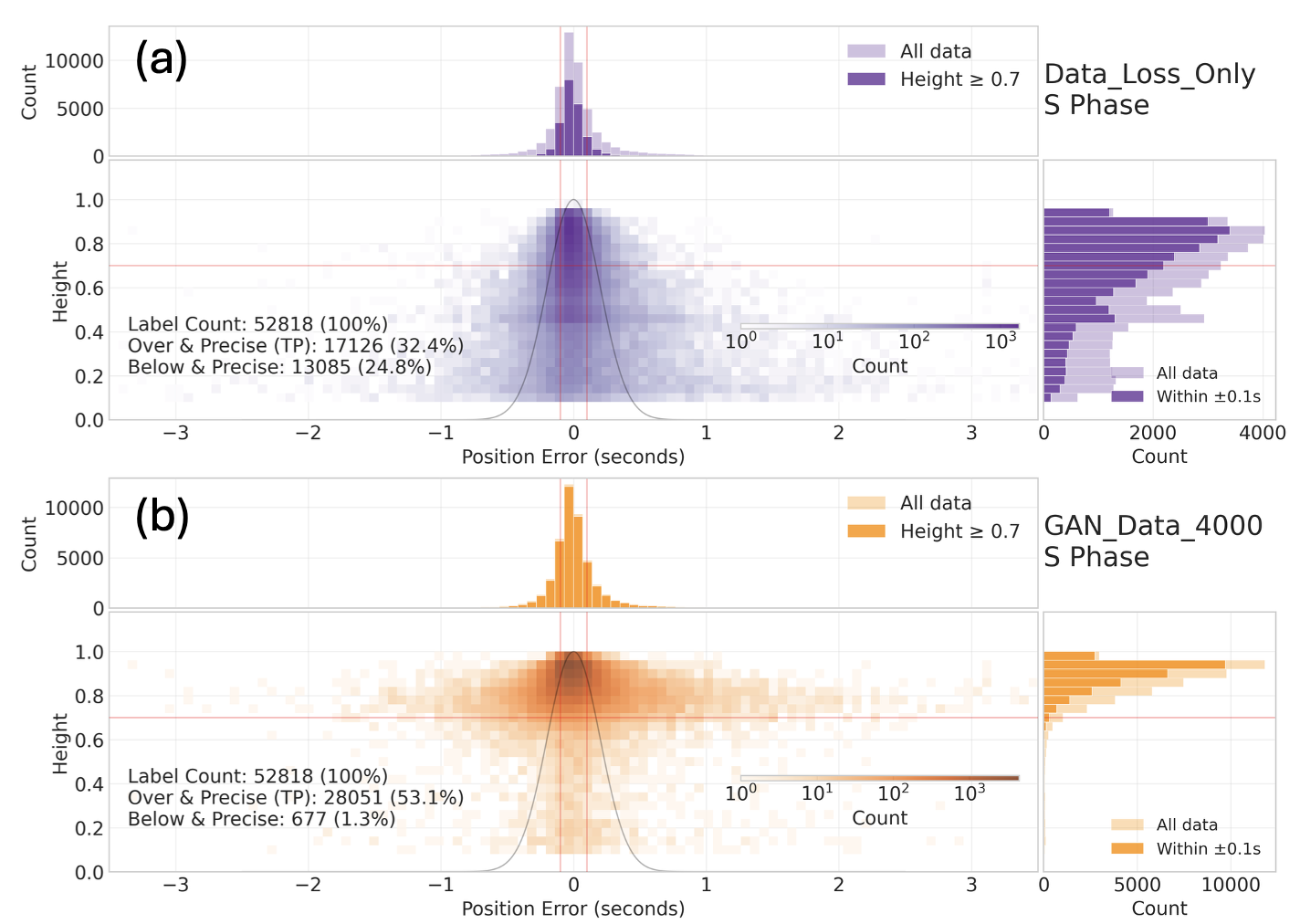}
\caption{\textbf{Large-scale statistical confirmation of amplitude suppression and its correction.}
Each point represents one S-phase prediction matched to a labeled arrival in the test set; x-axis shows temporal error from label center, y-axis shows peak amplitude. Black Gaussian curves represent reference labels. (a) Conventional method: dense horizontal suppression band at 0.5 amplitude confirms widespread amplitude suppression. Many predictions within this band are temporally accurate (within ±0.1 s, red lines) but ineffective due to insufficient amplitude. (b) Our framework: suppression band eliminated as the distribution is compressed above 0.7, with horizontal spread preserved. This recovers previously sub-threshold accurate predictions, achieving 64\% increase in effective S-phase recall (peak $>0.7$, time error $<0.1$ s).\\[6pt]
\textbf{Alt-text:} Two vertically stacked panels, conventional method on top and proposed method below. Each panel shows L-shaped arrangement with central 2D histogram and 1D marginal histograms above and to the right.}
    \label{fig:s_distribution}
\end{figure}

The suppression manifests as two distinct symptoms in the final prediction. The probability peak is suppressed at an amplitude of approximately 0.5, well below the 0.7 detection threshold, and its location is systematically displaced toward the later, high-amplitude boundary (Fig. \ref{fig:compare_runs}a, b). A large-scale statistical analysis of the entire test set (Fig. \ref{fig:s_distribution}a) confirms that a dense cluster of S-phase predictions hovers in a horizontal band around the 0.5 amplitude level. While many of these predictions are temporally accurate (within ±0.1s of the label), their insufficient peak height renders them ineffective, establishing suppression as a primary cause of detection failure.

With these empirical observations established, we now turn to diagnosing the underlying mechanism.

\section{Diagnosing the Mechanism}

\subsection{Three Interacting Factors}

The training history (Fig. \ref{fig:amplitude_suppression}) reveals a critical divergence. P-wave predictions develop successfully while S-wave predictions become trapped. Both phases initially evolve similarly, but their trajectories diverge as features differentiate. This contrast is diagnostic, and by identifying what differs between P and S, we can isolate the factors driving suppression. Our analysis reveals three interacting factors, none of which alone would create such a persistent trap.

\begin{figure}
    \centering
    \includegraphics[width=1\linewidth]{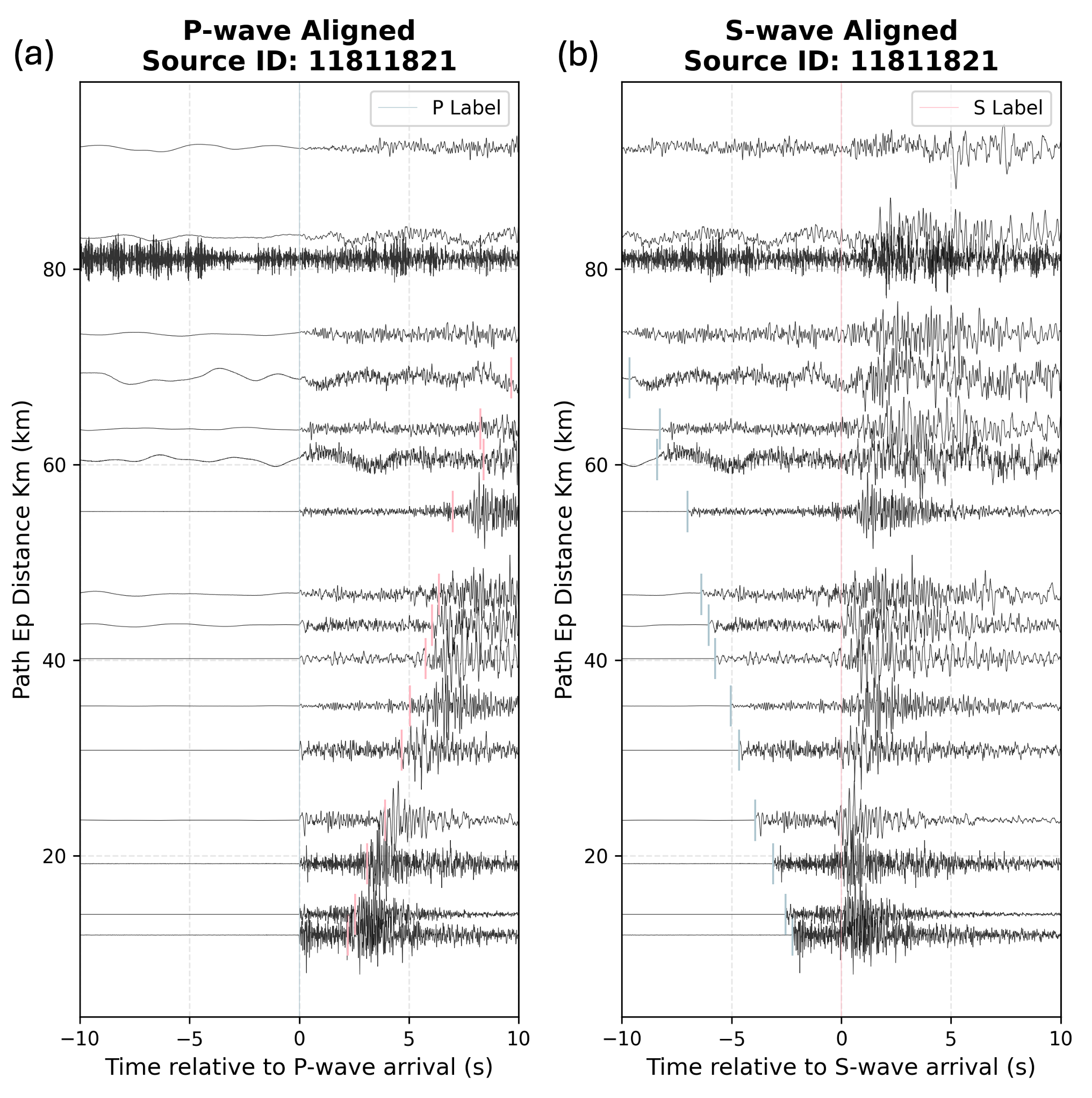}
\caption{\textbf{Comparative Analysis of P-wave and S-wave Arrival Time Labels.}
(a) P-wave: The amplitude boundary coincides with true arrival time, providing a distinct marker with low temporal uncertainty. (b) S-wave: Increasing epicentral distance causes waveform broadening and blurred amplitude boundaries. Longer propagation paths produce multiple overlapping waves, reducing amplitude gradients and annotation accuracy. This produces broad temporal uncertainty in S-wave arrivals at greater distances.\\[6pt]
\textbf{Alt-text:} Two side-by-side panels showing multiple z-channel waveforms from the same earthquake event, aligned by arrival labels. Left panel shows P-waves with sharp, well-aligned amplitude boundaries. Right panel shows S-waves with staggered, uneven amplitude boundaries.}
    \label{fig:compare_p_s}
\end{figure}

\textbf{Factor 1: Temporal uncertainty in S-wave arrivals.} During training, S-wave peaks oscillate dramatically left and right while growing (Fig. \ref{fig:amplitude_suppression}), whereas P-wave peaks ascend steadily with minimal lateral movement. This contrast suggests that S-wave labels vary widely in temporal position across training samples. The physical basis is clear: unlike P-waves where onset is defined by the amplitude boundary, S-wave onsets become increasingly ambiguous at greater epicentral distances. Dispersion separates S-waves from surface waves, signal attenuation degrades clarity, and manual picking accuracy suffers accordingly (Fig. \ref{fig:compare_p_s}). Because waveform characteristics differ substantially between near and distant events, CNN training effectively separates them into distinct groups. Temporal inaccuracies in public seismic datasets have been independently noted \citep{suarez2025pervasive-label}, and distant events, with fewer samples and inherently ambiguous onsets, become disproportionately susceptible.

\textbf{Factor 2: CNN correlation bias.} The evolutionary history of training reveals the source of this bias. The model first learns to identify noise by detecting sharp amplitude boundaries (Fig. \ref{fig:amplitude_suppression} step 235). Since P and S phases are defined as the complement of the Noise channel, they initially evolve together, sharing a common template anchored to these prominent boundaries. When features begin to differentiate, P-waves successfully detach because their labeled onsets coincide with amplitude boundaries. S-waves cannot detach because a gap exists between the boundary-anchored prediction and the labeled onset position. The S-phase prediction thus retains the original anchoring, and this trapped state exposes the model's underlying preference for strong amplitude gradients.

\textbf{Factor 3: Pointwise loss lacks lateral guidance.} In late training, S-wave predictions exhibit a distinctive oscillatory pattern where amplitude rises and falls cyclically without sustained growth. This vertical cycling with no horizontal correction suggests the loss function provides no lateral force. Binary Cross-Entropy treats each output position independently, and each vertical slice optimizes in isolation, providing only amplitude gradients without any mechanism to shift predictions horizontally. We confirm this geometric property through loss landscape visualization below.

\subsection{Loss Landscape Analysis}

\begin{figure}
    \centering
    \includegraphics[width=1\linewidth]{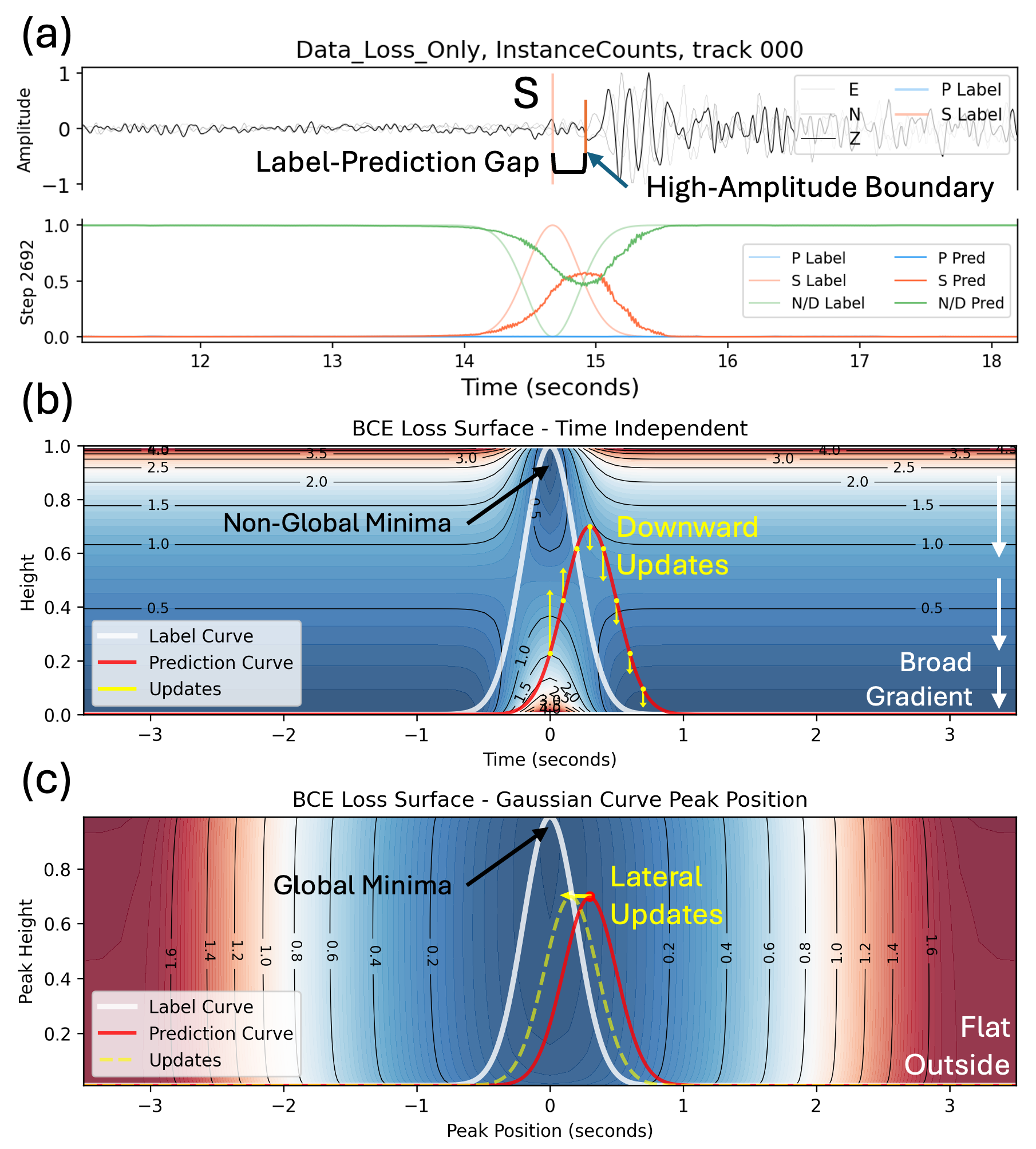}
\caption{\textbf{The optimization trap: Actual appearance, geometric origin, and conceptual solution.}
(a) Magnified view of a suppressed S-phase prediction showing characteristic flattened-top appearance. (b) Pointwise BCE loss landscape, constructed by stacking independent 1D loss curves along the time axis. Each vertical slice optimizes independently, providing only vertical gradients. Low-loss regions appear both at the label peak and on both sides when amplitude approaches zero, creating a non-global minimum. (c) Shape-constrained loss landscape, computed as the total BCE loss of a Gaussian template as its peak position traverses the amplitude-time plane. The holistic evaluation eliminates side minima, leaving a single global minimum with strong lateral guidance.\\[6pt]
\textbf{Alt-text:} Three vertically stacked panels. Top panel shows probability curve with flattened top. Middle panel shows 2D loss surface with steep downward slopes on both sides and upward peak only at center. Bottom panel shows bowl-shaped surface with minimum at center, flattening beyond 3 seconds.}
    \label{fig:landscape}
\end{figure}

We examined the geometric properties of the training objective to understand how these factors interact. A magnified view of a suppressed S-phase prediction (Fig. \ref{fig:landscape}a) reveals its characteristic flattened-top appearance. The theoretical loss landscape of pointwise BCE loss (Fig. \ref{fig:landscape}b) explains this geometry. Since pointwise loss treats each temporal position independently, the 2D surface consists of stacked independent 1D loss curves. Though each slice is independent, the 2D visualization becomes meaningful because temporal uncertainty causes label positions to vary across training samples, shifting the entire loss landscape laterally from step to step. This construction reveals the origin of the trap: low-loss regions appear both at the label peak and on both sides when amplitude is near zero. When predictions deviate temporally from the label center, the gradient points toward lower amplitude rather than toward the correct temporal position. No lateral force exists because adjacent positions are optimized independently, and the prediction becomes trapped in a state where amplitude suppression is geometrically enforced.

\subsection{Numerical Validation}

\begin{figure*}
    \centering
    \includegraphics[width=1\linewidth]{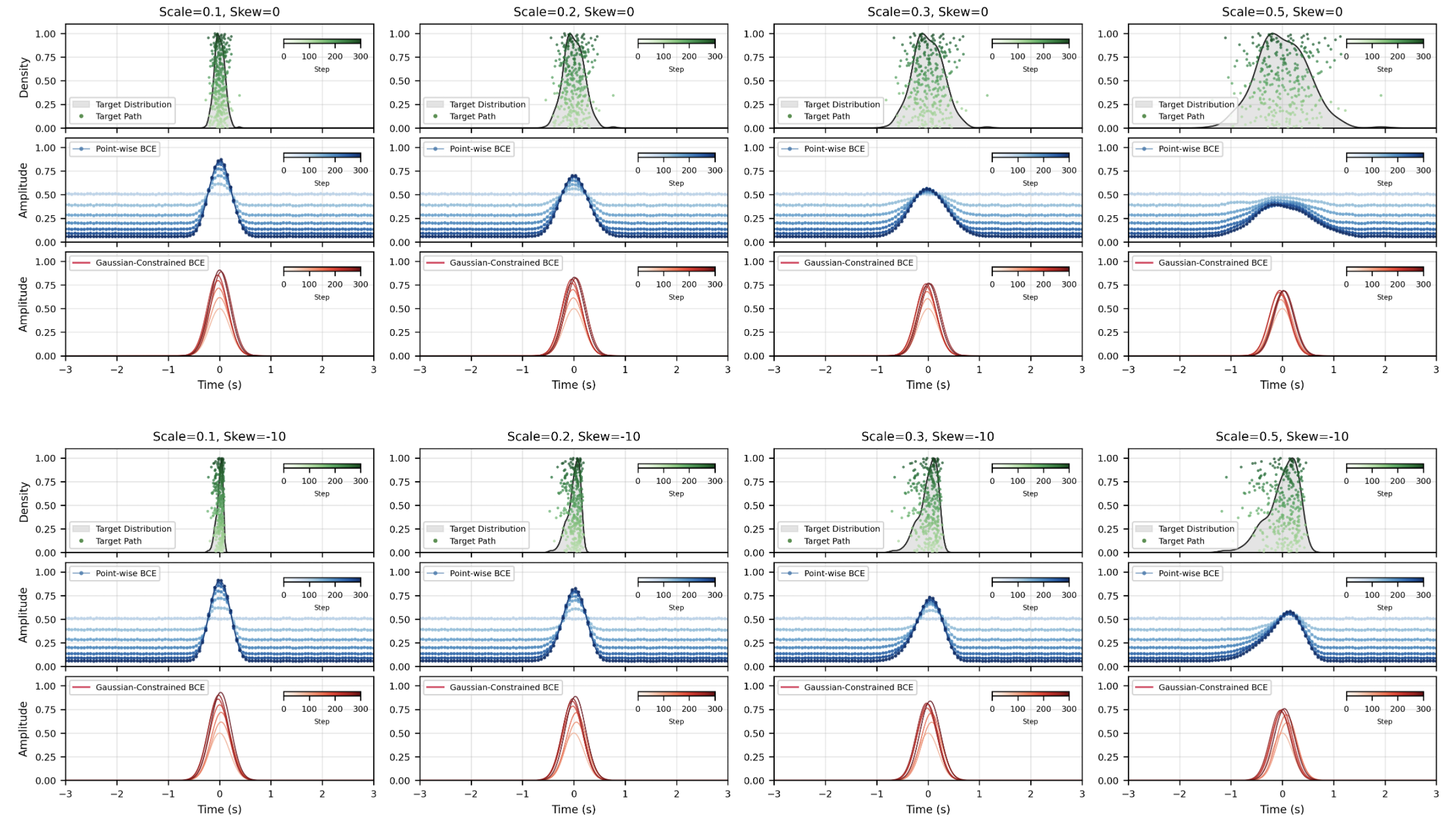}
\caption{\textbf{Numerical validation: Deconstructing the key factors of the optimization trap.}
This experiment removes the neural network entirely: trainable parameters represent prediction values directly, updated via PyTorch gradient descent on BCE loss. The 2×4 grid varies distribution width ($\sigma$ = 0.1, 0.2, 0.3, 0.5, columns) representing temporal uncertainty and skewness (0 vs. -10, rows) representing systematic bias. At each training step, a target position is sampled from the specified distribution. Each subplot shows three panels: green (target sampling distribution), blue (pointwise optimizer output), red (Gaussian-constrained optimizer output). Results demonstrate temporal uncertainty controls amplitude suppression; systematic bias introduces temporal offset only when combined with high uncertainty.\\[6pt]
\textbf{Alt-text:} Grid of 8 subplots arranged in 2 rows by 4 columns. Each subplot shows three overlaid curves in green, blue, and red. Blue curves progressively flatten from left to right columns; red curves maintain peak shape throughout.}
    \label{fig:simulation}
\end{figure*}

To isolate how temporal uncertainty and systematic bias individually contribute to suppression, we designed a numerical simulation that removes the neural network entirely (Fig. \ref{fig:simulation}). Trainable prediction values are directly optimized by gradient descent on the theoretical loss landscape, with targets sampled from distributions of varying width (temporal uncertainty) and skewness (systematic bias).

The simulation reveals a clear pattern. Temporal uncertainty controls amplitude: as the target distribution widens, the optimizer repeatedly enters and exits the label's core region, preventing amplitude from growing consistently. High temporal uncertainty alone induces amplitude suppression. Systematic bias requires uncertainty to cause offset: when uncertainty is low, the optimizer achieves high amplitude despite bias, but combined with high temporal uncertainty, systematic bias causes significant temporal displacement. The bottom-right profile of the pointwise optimizer closely resembles the observed suppression pattern, confirming this simulation captures the trap's core mechanism.

The pointwise optimizer is not failing but rather succeeding at the wrong objective. The flattened output closely matches the expected value of all possible label positions under temporal uncertainty. This is a valid probabilistic answer, but not the Gaussian shape we seek. The core issue is that our task requires the peak to shift to the correct temporal position, but pointwise optimization prevents neighboring positions from coordinating. When one position receives a gradient to decrease amplitude, adjacent positions receive no corresponding signal to increase. The peak cannot move laterally because each position optimizes in isolation. This mismatch reveals why the solution requires redefining the optimization objective: shifting from pointwise probability to shape-aware optimization.

\section{The Shape-Then-Align Strategy}

The diagnosis points to a fundamental mismatch between how labels are designed and how they are optimized. Phase arrival labels are Gaussian templates \citep{zhu2018phasenet-82a, park2025divide} where each point's value depends on its position relative to the arrival onset. Yet pointwise BCE treats these templates as independent probability estimates, fragmenting the coherent structure that defines them.

This mismatch suggests the solution: shape-aware training objectives must preserve structural coherence rather than fragment it. We call this the shape-then-align strategy. The geometric template must stabilize first because shape variations produce far larger loss changes than small temporal shifts. When shape remains unstable, even a correctly aligned prediction generates no appreciable loss improvement, as the shape-induced loss dominates. Temporal alignment can only succeed once the template is coherent enough for positional refinements to register in the loss landscape.

The loss landscape supports this intuition. When we constrain the model to output only a perfect Gaussian template (Fig. \ref{fig:landscape}c), the resulting surface reveals a single, well-defined global minimum at the correct time and full amplitude. The numerical validation (Fig. \ref{fig:simulation}) confirms this prediction: an optimizer constrained to Gaussian-shaped outputs maintains substantially higher peak amplitudes while remaining anchored at the center of the target distribution, avoiding the drift observed in pointwise optimization.

The strategy is validated. The implementation challenge is enforcing shape coherence in practice while retaining flexibility.

\section{cGAN Implementation}

\begin{figure}
    \centering
    \includegraphics[width=0.8\linewidth]{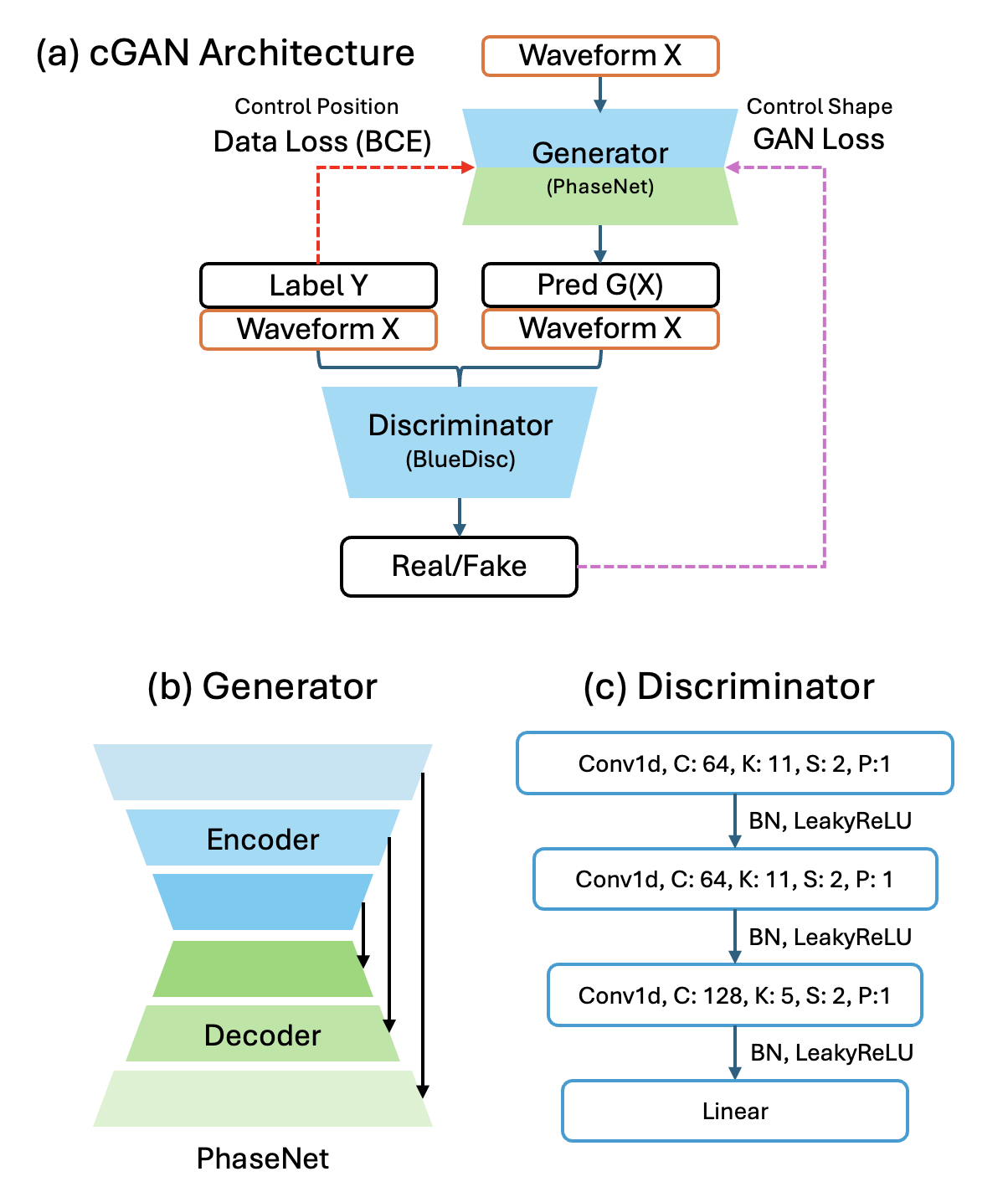}
\caption{\textbf{Model Architecture.}
(a) Complete cGAN training framework: BCE-based Data Loss controls temporal position, GAN Loss from discriminator controls shape. (b) Generator: PhaseNet (U-Net encoder-decoder). (c) Discriminator (BlueDisc): lightweight CNN with three blocks.\\[6pt]
\textbf{Alt-text:} Three-part diagram. Part a shows flowchart with generator in center receiving two loss signals: data loss from dataset labels and GAN loss from discriminator judgment. Part b shows generator as hourglass-shaped encoder-decoder. Part c shows discriminator as funnel-shaped stack, wide at top and narrowing to a single Linear output at bottom.}
    \label{fig:model_architecture}
\end{figure}

Among various shape-aware approaches, we chose conditional GANs (cGAN; \citealt{mirza2014conditional-023}), building on the generative adversarial framework \citep{goodfellow2020generative}, for this proof of concept. A learned discriminator can enforce shape constraints without explicit template definitions. It autonomously discovers valid Gaussian-like outputs by distinguishing real labels from generated predictions. The generator is a standard PhaseNet model \citep{zhu2018phasenet-82a}, using the implementation provided in SeisBench \citep{woollam2022seisbencha-bde}, and we add only a lightweight CNN discriminator (BlueDisc) conditioned on the input waveform (Fig. \ref{fig:model_architecture}).

The training objective combines adversarial and data-fidelity components. The adversarial loss encodes a minimax game where the discriminator $D$ learns to distinguish real labels from generated ones, while generator $G$ learns to fool it:
$$\mathcal{L}_{\text{cGAN}} = \mathbb{E}[\log D(y, x)] + \mathbb{E}[\log(1 - D(G(x), x))]$$
The BCE loss enforces instance-wise pairing between each waveform and its ground-truth label:
$$\mathcal{L}_{\text{BCE}}(y, \hat{y}) = -\frac{1}{N} \sum_{i=1}^{N} [y_i \log(\hat{y}_i) + (1-y_i) \log(1-\hat{y}_i)]$$
The final objective combines both:
$$\mathcal{L}(G, D) = \mathcal{L}_{\text{cGAN}} + \lambda \mathcal{L}_{\text{BCE}}$$
Following the pix2pix framework \citep{isola2017image}, we combine adversarial loss with a data-fidelity term, substituting L1 with the BCE loss originally used in PhaseNet training, as both serve to align predictions with ground-truth labels. The weight $\lambda = 4000$ balances temporal anchoring (BCE) against shape coherence (GAN). We trained on the INSTANCE dataset \citep{michelini2021instance-b0c}, specifically the InstanceCounts version, a deliberately unfiltered compilation preserving real-world imperfections essential for exposing temporal uncertainty. Training used 699,980 samples, 10,000 steps, batch size 100, learning rate $10^{-3}$ \citep{mnchmeyer2022which-305}, Adam optimizer \citep{kingma2014adam-6be} with $\beta_1 = 0$ \citep{huang2024gan}, and random seed 42 on an NVIDIA RTX 3090 for reproducibility. Full implementation is available at our code repository.

\begin{figure*}
    \centering
    \includegraphics[width=1\linewidth]{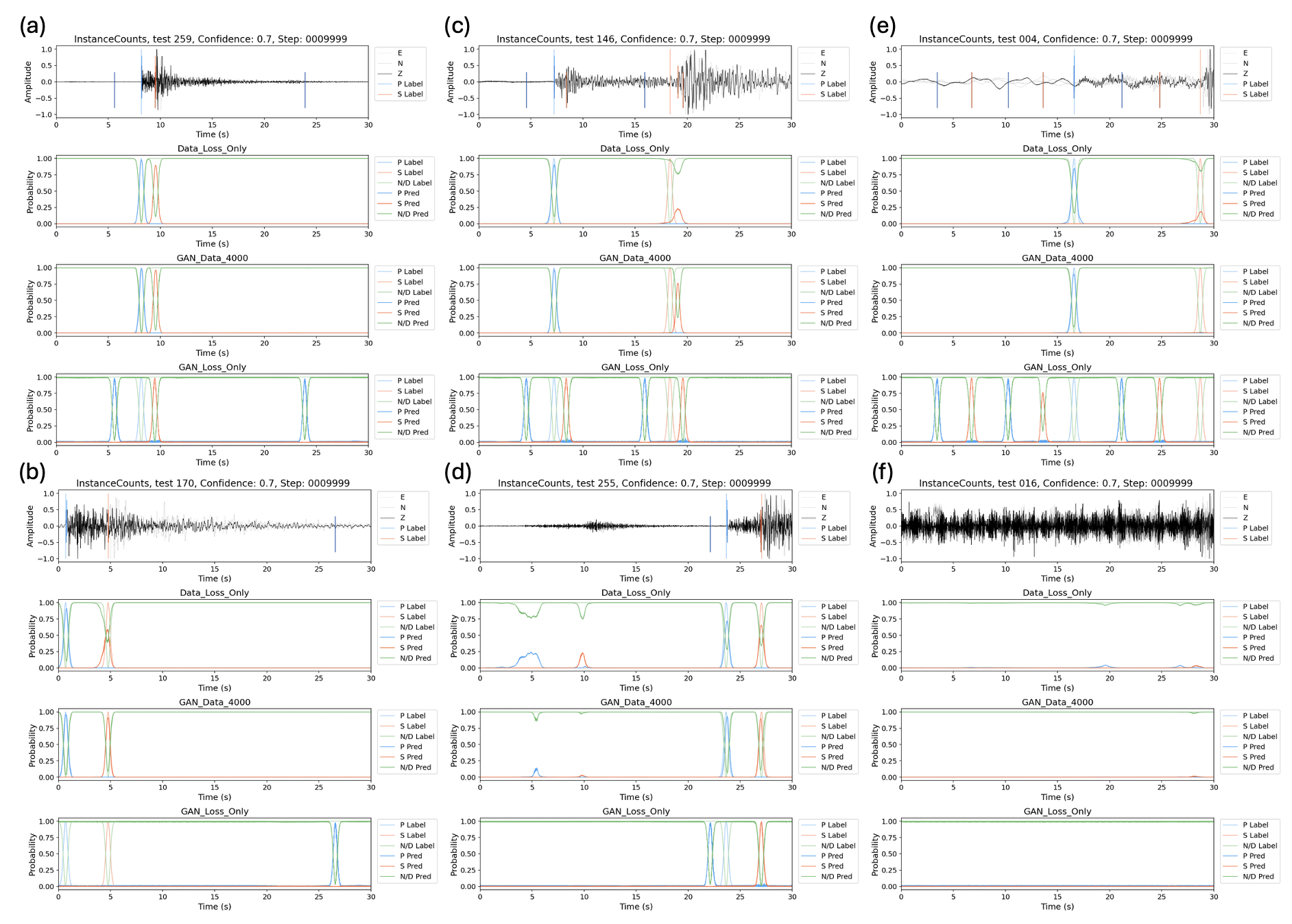}
\caption{\textbf{Diverse case analysis.}
Hybrid loss shows robustness: near-field earthquakes (a, b) and noise (f). Limitations: regional event (c) shows miscalibrated timing; double-earthquake (d) misses smaller signal. Pure GAN (bottom row of each panel) produces stereotyped predictions across all examples regardless of input.\\[6pt]
\textbf{Alt-text:} Grid of 6 panels labeled a through f. Each panel shows 3-channel seismic waveform at top with three rows of probability predictions below, comparing conventional, hybrid, and pure GAN methods.}
    \label{fig:examples}
\end{figure*}

The results confirm the shape-aware strategy works. The statistical distribution (Fig. \ref{fig:s_distribution}b) shows complete elimination of the suppression band at 0.5 amplitude, elevating temporally accurate predictions above the detection threshold. This recovers previously sub-threshold arrivals, achieving a 64\% increase in effective detections (peak $>0.7$, time error $<0.1$ s). Analysis across diverse scenarios (Fig. \ref{fig:examples}) reveals robustness in common cases (near-field earthquakes, noise records) but limitations in extreme cases (regional events, double earthquakes). Pure GAN training produces stereotyped predictions, confirming BCE's necessity as a stabilizing anchor.

\begin{figure}
    \centering
    \includegraphics[width=0.9\linewidth]{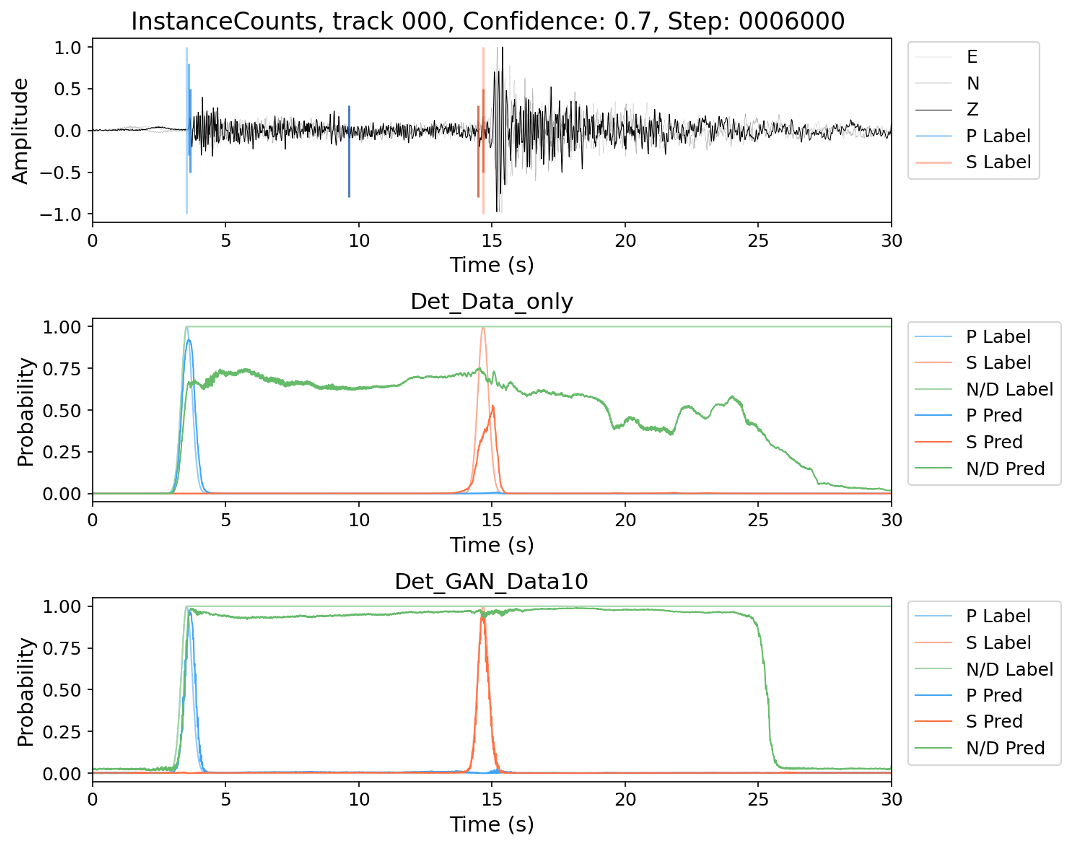}
\caption{\textbf{Generality as a shape learner.}
Training with tapered boxcar labels using the same architecture demonstrates that the discriminator learns adaptive geometry beyond the specific Gaussian target. See Video 3 in Data and Resources for the training animation.\\[6pt]
\textbf{Alt-text:} Single-event visualization with 3-channel waveform at top and probability predictions below. Predictions show rectangular shapes with tapered edges matching the boxcar-style labels.}
    \label{fig:det_compare_run}
\end{figure}

A key strength of the adversarial approach is adaptivity. Training on modified labels such as tapered boxcar functions instead of Gaussians demonstrates the discriminator learns abstract shape concepts, not just fixed templates (Fig. \ref{fig:det_compare_run}).

\section{Discussion}

One practical consideration is that breaking amplitude suppression may increase detection counts while decreasing precision when measured against conventional catalogs. This apparent trade-off reflects a change in detection philosophy rather than a degradation in performance. By revealing peaks that were previously hidden below detection thresholds, we expose signals that conventional methods categorically excluded (Fig. \ref{fig:s_distribution}). These newly surfaced detections can then be filtered through downstream processing such as association and location \citep{sun2023deep-chihshang}. The goal shifts from suppressing uncertain signals to exposing them for informed decision-making.

The diagnostic framework we introduce extends beyond the specific cGAN implementation. Any shape-aware loss function should break amplitude suppression by providing the lateral guidance absent in pointwise optimization. Overlap-based losses like Dice, as applied to phase picking by \citet{park2025reducing-831}, evaluate predictions holistically rather than pointwise, inherently coupling neighboring positions. Probabilistic regression approaches \citep{williams2023deep} that model arrival time distributions sidestep the mismatch between structured templates and pointwise optimization entirely. These independent developments converge on the same insight: preserving coherence across the output structure.

More broadly, the loss landscape visualization and numerical simulation techniques provide a general methodology for analyzing label design choices. From simple variations in Gaussian width to complex multi-component templates like those explored by \citet{park2025divide}, the geometric properties of any label shape can be systematically evaluated before training. This predictive capacity transforms label design from empirical trial-and-error into principled analysis.

The training dynamics reveal additional insights worth exploring. The third output channel, beyond P and S, plays an underappreciated structural role. Because it is defined as the complement of P and S, it forces the model to structure the entire waveform rather than treat phases as isolated events. The model learns holistic waveform structure through this channel first, with P and S emerging as derivatives of this shared representation. Different label designs produce markedly different evolutionary trajectories. In conventional Gaussian labels (Fig. \ref{fig:compare_runs}), this channel represents Noise \citep{zhu2018phasenet-82a}, and S-waves anchor to amplitude boundaries (see Video 1 in Data and Resources). In boxcar-style Detection labels (Fig. \ref{fig:det_compare_run}), this channel represents the earthquake detection window  \citep{mousavi2020earthquake-3d2}, and training dynamics change substantially (see Video 3 in Data and Resources). These observations suggest the third channel provides higher-order structural constraints that deserve further investigation.

\section{Conclusion}

Amplitude suppression is not a mysterious performance bottleneck but a predictable consequence of how pointwise loss interacts with temporal uncertainty. Three factors combine to create an optimization trap. Temporal uncertainty in S-wave arrivals causes labels to shift across training samples. CNN correlation bias anchors predictions to sharp amplitude boundaries rather than subtle onsets. Pointwise BCE loss provides only vertical gradients, lacking the lateral forces needed to correct temporal misalignment. Conventional training cannot escape this trap. The shape-then-align strategy we propose addresses these factors by enforcing structural coherence before optimizing temporal position. Our cGAN implementation recovers sub-threshold S-wave arrivals, achieving a 64\% increase in effective detections on the INSTANCE dataset.

The diagnostic framework extends beyond this specific implementation, enabling informed choices among alternative shape-aware approaches. The loss landscape visualization and numerical simulation techniques we introduce predict how label designs will interact with data characteristics before training. This transforms label design from trial-and-error into principled analysis. Understanding loss geometry is as critical as understanding model architecture.

\section*{Data and Resources}
The INSTANCE dataset, specifically the InstanceCounts version, is available at \href{https://www.pi.ingv.it/banche-dati/instance/}{https://www.pi.ingv.it/banche-dati/instance/}. Training dynamics are visualized in supplementary videos: Videos 1 and 2 show conventional PhaseNet training on Gaussian labels for 10,000 steps (\href{https://youtu.be/rdnCe1mJpA8}{https://youtu.be/rdnCe1mJpA8}) and 100,000 steps (\href{https://youtu.be/tNiBvxP69uM}{https://youtu.be/tNiBvxP69uM}), respectively. Video 2 is an extended version demonstrating that amplitude suppression persists even with prolonged training. Video 3 shows training with tapered boxcar labels (\href{https://youtu.be/DCB8SHBMJEg}{https://youtu.be/DCB8SHBMJEg}). Our implementation builds upon SeisBench \citep{woollam2022seisbencha-bde} (\href{https://github.com/seisbench/seisbench}{https://github.com/seisbench/seisbench}) and pick-benchmark \citep{mnchmeyer2022which-305} (\href{https://github.com/seisbench/pick-benchmark}{https://github.com/seisbench/pick-benchmark}). Core software dependencies include PyTorch (\href{https://pytorch.org}{https://pytorch.org}), ObsPy (\href{https://obspy.org}{https://obspy.org}), and Matplotlib (\href{https://matplotlib.org}{https://matplotlib.org}). The complete implementation, including model architecture, training scripts, numerical simulation code, and hyperparameter configurations, is available at \href{https://github.com/SeisBlue/BlueDisc}{https://github.com/SeisBlue/BlueDisc}. AI language models (Claude, Anthropic; Gemini, Google) assisted with specific coding tasks including framework migration and visualization scripts; all methodology, experimental design, and scientific conclusions originated from the authors, and all code and results have been verified. All URLs last accessed January 2026.

\section*{Declaration of Competing Interests}
The authors declare no competing interests.

\section{Acknowledgments}
We thank the editor and anonymous reviewers for their constructive comments that improved this manuscript.


\end{document}